\documentclass[onecolumn]{IEEEtran}
\IEEEoverridecommandlockouts
\usepackage[final]{graphicx}
\usepackage{srcltx}
\usepackage{cite}
\usepackage{upgreek}
 \usepackage{subfigure}
 \usepackage{subfigmat}
\usepackage{amssymb}
\usepackage{amsmath}
\usepackage{amsthm}
\usepackage{amsfonts}
\usepackage{bm}
\usepackage{makecell}
\usepackage{xcolor}
\usepackage{balance}
\usepackage{float}
\usepackage{algorithm}
\usepackage{algpseudocode}
\usepackage{algorithmicx}
\usepackage{cleveref}
\graphicspath{{./Figs/crop/}}

\begin{document}

\title{Code Design for Non-Coherent Detection of Frame Headers in Precoded Satellite Systems}

\author{Farbod Kayhan and Guido Montorsi
\thanks{Farbod Kayhan is with the Interdisciplinary Centre  for  Security, Reliability  and  Trust (SnT), University of Luxembourg (email: farbod.kayhan@uni.lu). Guido Montorsi is with the Department of Electronics and Telecommunication, Politecnico di Torino (email: guido.montorsi@polito.it.}}

\newtheorem{Theorem}{Theorem}[section]
\newtheorem{Lemma}{Lemma}[section]
\newtheorem{Definition}{Definition}[section]
\newtheorem{Conjecture}{Conjecture}[section]
\newtheorem{Corollary}{Corollary}[section]
\newtheorem{Proposition}{Proposition}[section]

\newcommand{\defeq} {\overset{\underset{\mathrm{def}}{}}{=}}
\newcommand{\De} {\mathrm{\bm{\Delta}}}
\newcommand{\Tht} {\mathrm{\bm{\Theta}}}
\newcommand{\g} {\mathrm{\bm{\gamma}}}
\newcommand{\HH} {\mathrm{\bf{H}}}
\newcommand{\bbb}{\mathrm{\bf{b}}}
\newcommand{\I}{\mathrm{\bf{I}}}
\newcommand{\A}{\mathrm{\bf{A}}}
\newcommand{\aaa}{\mathrm{\bf{a}}}
\newcommand{\h}{\mathrm{\bf{h}}}
\newcommand{\s}{\mathrm{\bf{s}}}
\newcommand{\x}{\mathrm{\bf{x}}}
\newcommand{\uu}{\mathrm{\bf{u}}}
\newcommand{\C}{\mathbb{C}}
\newcommand{\D}{\mathcal{D}}
\newcommand{\SPSK}{\text{-S}\text{PSK}}
\newcommand{\SQPSK}{\text{-S}\text{QPSK}}
\newcommand{\SAPSK}{\text{-S}\text{APSK}}
\newcommand{\SkPSK}{\text{-S}^k\text{PSK}}
\newcommand{\SkQPSK}{\text{-S}^k\text{QPSK}}
\newcommand{\SkAPSK}{\text{-S}^k\text{APSK}}
\newcommand{\PSNR}{\text{PSNR}}
\newcommand{\SNR}{\text{SNR}}
\newcommand{\LLR}{\text{LLR}}

\maketitle

\begin{abstract}
In this paper we propose a simple method for generating short-length rate-compatible codes over
$\mathbb{Z}_M$ that are robust to non-coherent detection for $M$-PSK constellations. First, a
greedy algorithm is used to construct a family of rotationally invariant codes for a given
constellation. Then, by properly modifying such codes we obtain codes that are robust to
non-coherent detection. We briefly discuss the optimality of the constructed codes for special
cases of BPSK and QPSK constellations. Our method provides an upper bound for the length of optimal
codes with a given desired non-coherent distance. We also derive a simple asymptotic upper bound on
the frame  error rate (FER) of such codes and provide the simulation results for a
selected set of proposed codes. Finally, we briefly  discuss the problem of designing binary codes that are robust to
non-coherent detection for QPSK constellation.

\end{abstract}


\IEEEpeerreviewmaketitle

\section{Introduction}\label{sec:introduction}
In this paper we address the problem of short-length rate-compatible (RC) code design for
non-coherent channels. In order to motivate the subject, we describe a specific system scenario,
namely signaling in a precoded satellite system, where short codes robust to non-coherent detection
are required by the system model for physical layer header (PLH). As we will see, in such systems, the phases of the received symbols are changed by an unknown random value which remains constant during the transmission of each codeword. Such a scenario can be considered as a special case of a more general model where the phases are changed at the symbol level following a stochastic process with given parameters.
Both coding and modulation design for non-coherent channels have been addressed since very first
days of communication theory (see for example \cite{scholtz} and references therein)
and hence the literature on the subject is quite rich, addressing varieties of system models and
strategies.

Two main approaches are presented in the literature following essentially two different strategies.
The first approach is based on pairing a capacity achieving code (LDPC, Turbo, ...) with an
optimized constellation space for non-coherent channel. In this approach, the main goal is to find
the capacity achieving (optimal) distribution of the signal space for the given channel model. For
details on some of the results obtained in this direction we refer the readers to
\cite{colavolpe_capa,katz,Nuriyev}. It is important to notice that in all these works the
constellation is considered to be infinite and the system performance is studied asymptotically.
For finite constellation sets, the problem of finding the optimal constellation is in general an
open problem. The second strategy is to fix the constellation shape and then trying to design the
codebook (and possibly the constellation labeling) for non-coherent detection. This is also the
strategy that we pursue in this paper. The main idea in this case is to maximize the so called
\emph{non-coherent equivalent distance} of the code\footnote{Notice that the non-coherent distance has been defined in several ways in the literature.
In this paper we adopt the definition which is in line with that of \cite{ASMSerror}.}. The coding design problem for specific cases of
M-PSK, APSK, and QAM constellations has been investigated by several authors, see for example
\cite{Knopp,Sun_Leib,Wei_Lin,Wei_Chen,Wei,richardson} and references within. In the following, we
briefly review some of the above works that are directly related to the problem we are considering
and mention their differences compared to our approach.

In \cite{Knopp} the authors study the coding design problem for the non-coherent AWGN channel
assuming a $M$-ary phase shift keying (PSK) modulation scheme. In particular, a coding design over
$\mathbb{Z}_M$ based on exclusion of unwanted codewords is proposed and codes up to length $N=15$
and information vector size $K=10$ over $\mathbb{Z}_4$ and $\mathbb{Z}_8$ are constructed. In this
method there exist an isomorphism between the ring over which the code is defined and the
modulation scheme. This isomorphism also implies the constellation labeling. The results in
\cite{Knopp} are mainly based on numerical approaches. F-W. Sun and H. Leib in \cite{Sun_Leib}
extend these results by providing a analytical framework. This is done by showing the relation
between the non-coherent distance and those of Euclidean and Lee distances. In a series of papers,
the authors in \cite{Wei_Lin, Wei_Chen,Wei} study the block-coded modulation for MPSK, QAM and APSK
constellations. In particular, the minimum non-coherent distance is obtained separately for each
constellation and a given labeling.

The design strategies introduced in previously mentioned works do not perfectly match with the
system scenario that we are interested in this paper.  Indeed, to the best of our knowledge, no
study on the optimal code lengths as a function of $K$ and minimum distance have been presented in the
literature. In this paper, we first discuss in some details the system model that we are interested
in and motivate the reasons for which short-length codes with non-coherent detection are needed in
such systems. Then, we propose a method to obtain such codes by modifying the generator matrix of
rotationally invariant codes. A greedy algorithm is used to optimize the minimum distance of rotationally invariant codes with
codeword lengths up to $N=256$ and information vector lengths up to $K=15$. The results are
reported only for BPSK and QPSK constellations. However, the method can be
extended for any $M$-PSK constellation. These results provide a lower bound on the minimum distance of rotationally invariant codes
for a given pair $[N,K]$. Finally, we derive a simple asymptotic upper bound on the frame error rate (FER) of the
designed codes. This will help us to choose a code from the table with desired FER for the PLH of
the system.

The rest of this paper is organized as follows. In Section \ref{sec:sysmodel}, we describe our
system model and define the problem we intend to solve. In Section \ref{sec:construct}, we present the proposed design
for the rotationally invariant codes and provide the tables for the optimized codes for information lengths up to $K=14$.  In Section \ref{sec:upperbound} we derive a simple
upper bound of the error probability as a function of SNR and the minimum non-coherent equivalent distance of
a code. We also explain the strategy to build such a codes using the rotationally invariant codes. This upper bound is then used to select the desired codes to design the PLH in Section \ref{sec:simul}, where we also
present the simulation results for a selected set of codes. Finally, we conclude
the paper in Section \ref{sec:conc} and discuss some possible future research directions.

\section{System Model and Problem Definition}\label{sec:sysmodel}
Non-coherent detection is an attractive technique in several communication scenarios. Two examples
are systems without carrier phase tracking and AWGN channels with flat fading channels when the
effects of the phase rotation is considered independently of the amplitude variation. In this
paper, we consider a precoded satellite system and explain the reasons for which non-coherent
detection of PLH is needed in such system scenarios.

In order to describe our system model we confine ourselves to the specifics of the  digital video
broadcasting standard (DVB-S2X) \cite{DVBs2x}. However, our assumptions are valid for a wide range
of satellite communication systems (not broadcasting) where adaptive coding and modulation (ACM) is
employed. In ACM schemes the forward error correction (FEC) rate and the modulation schemes are
selected from a list depending on the channel state information (CSI) in order to maximize the
throughput of the channel \cite{ACM}. At the receiver side, before detecting/decoding the
FEC frame, one needs the knowledge about the coding rate and the modulation scheme. These
information, referred to as physical layer header (PLH), are encoded and sent before each FEC
frame. The PLH headers in DVB-S2X and DVB-T have constant length, however, as it has been shown in
\cite{ACMIconf,ACMI,patent} considerable gains on average code lengths may be obtained by employing
variable length coding technique for PLH.

In DVB-S2X, in order to track the carrier phase, usually a sequence of pilot symbols are
transmitted in between of blocks of data sequences. The pilot symbols are chosen from the QPSK
constellation and are known to the receiver. By using techniques such as phase-locked loop (PLL)
one can estimate the carrier phase for all data sequences. The residual phase noise after the PLL
is usually modeled as white noise and can be further handled by, for example, optimizing the
constellation space \cite{TWC}.

Recently, it has been shown that precoding can provide significant gains in multi-beam satellite
systems with interference \cite{vazquez,alessandro}. In this paper, we consider the effect of
precoding on the detection of PLH and FEC frame. The main problem is that the precoder matrix, used
to reduce interference, is known at transmitter but not at the receiver. The effect is that
received signal in the precoded section may be affected by a constant but unknown changes of the phase
(and the amplitude) of the received signal. In such cases, the common pilots, not precoded, cannot
be used to estimate the phase in the precoded parts. One way to solve this problem is to design the
PLH code such that non-coherent detection is possible. In this way, after PLH detection, we can
derive an estimation of the phase change due to the precoding matrix for the following FEC block of
data. The PLH codes are usually short, having lengths up to few hundreds of symbols (90 symbols for
normal MODCODs and 900 symbols for VL-SNR MODCODs in DVB-S2). One of the characteristics of such a
system, that distinguish it from the previous system models, is the fact that not all PLH in a
super frame may be precoded with the same precoder. Therefore, the phase estimation may be
independent from one PLH to another. Our main problem is then to design short finite-block length
codes for PLH that are robust to non-coherent detection. In the next section we provide a simple
way to construct such family of codes.

The received signal can be written as $y_k = s_k e^{j\theta_k} + n_k$  where $s_k$ is the
transmitted symbol, $n_k$ is the additive white Gaussian noise and $\theta_k$ is  modeled as a
random process with first order statistic uniform in $[0,2\pi]$. Different code design approaches
stem from the difference of the coherence time $T_\theta$ of the phase process $\theta_k$ and the
codeword length $N$. Given the above discussion, in our case, we assume $T_\theta\gg N$ so that we
can drop the subscript $k$ from $\theta$ and write $y_k = s_k e^{j\theta} + n_k$.

\section{Rate compatible Rotationally invariant codes}\label{sec:construct}
We start the section by presenting a simple and general greedy algorithm for constructing  families of RC
codes with small dimension $K$  over $\mathbb{Z}_2$.
We then extend the procedure to construct codes over arbitrary groups  $\mathbb{Z}_M$ and show how to particularize it
to construct rotationally invariant codes for $M$-PSK.
We also show that the designed code families are competitive with the optimal codes that are known in some cases.

\subsection{Construction of rate-compatible linear binary codes}\label{sec:algorithm}
In this part we provide a simple approach based on greedy algorithm to construct the generating
matrices $\mathbf{G}^{(n)}$ of a family of good rate-compatible codes over $\mathbb{Z}_2$ with small dimension $K$. The
step by step description of the algorithm is provided in Algorithm \ref{alg:1}. In step 2, we
initialize the generator matrix as $\mathbf{G}^{(0)}$ to the empty matrix.  In step 4, $w_2()$
denotes the Hamming weight of a binary vector.
The algorithm adds, at each step, a new column to the generator matrix such that the number of
nearest neighbor codewords is reduced as much as possible, and hence possibly increase the minimum
distance. The algorithm stops when a target desired minimum distance, $d_{min}^t$, or a maximum codeword size
is achieved. The optional condition in step 6 guarantees that when $n=2^K - 1$, the optimal
maximum length codes (all possible columns in generating matrix) are obtained.
Complexity of the algorithm is exponential with $K$  (step 4 and 5) but linear with the code length $n$.
\begin{algorithm}[!tp]
	\caption{Greedy Approach to Construct the Generating Matrix of a small dimension $K$ RC
Compatible Code}
	\label{alg:1}
	\begin{algorithmic}[1]
		\linespread{1.2}
        \State Input: target minimum distance $d_{min}^{t}$, $K$
        \State Initialize: Set  codeword length $n=0$, $d_{min}^{0}=0$ and $\mathbf{G}^{(0)}=()$.
  \State For all increasing $n$ starting from 0:
  \State Construct the list of input words $\mathcal{U}$ generating the minimum distance
      codewords (nearest neighbors):
      \[   \mathcal{U}=\left\{\mathbf{u}=\arg \min \big( w_2(\mathbf{u^TG}^{(n)}) \big) \right\}.\]
     \State Find a new column vector $\mathbf{g}^{(n+1)}$, to be appended to $\mathbf{G}^{(n)}$,  that reduces as
     much as possible the number of nearest neighbors
       \begin{equation}
      \mathbf{g}^{(n+1)}\in \arg\max_{\mathbf{g}}\sum_{\mathbf{u}\in \mathcal{U}} \mathbf{g^T\cdot u}\;\;
      \label{eq:opt},
      \end{equation}
      where ``$\cdot$'' denotes the scalar product (over $\mathbb{Z}_2$).
      If there is  more than one vector satisfying (\ref{eq:opt}), pick a random one.

\State Optionally one may also enforce that the new vector $\mathbf{g}^{(n+1)}$ is different from
all the  columns of  $\mathbf{G}^{(n)}$.
      \State Append the column  $\mathbf{g}^{(n+1)}$ to $\mathbf{G}^{(n)}$:
           \[
          \mathbf{G}^{(n+1)}=\left(\mathbf{G}^{(n)}|\mathbf{g}^{(n+1)}\right),
          \]
       \State $n=n+1$ and goto 4 if $d_{min}^{t}$ is not achieved.
	\end{algorithmic}
\end{algorithm}

Despite its simplicity, this algorithm provides families of RC codes with performances close to
those of optimal codes. Such a comparison is done in Table \ref{tab:codes1}. In this table we
report the required length needed to satisfy a target minimum distance for given $K$. The columns
labeled  {\textbf{B}} correspond to the optimal codes. The optimal minimum distances
are known for all values of $K=2,\ldots,15$ and $N=2,\ldots,250$. Routines to construct optimal
codes can be found in \cite{optcode}. The results for the codes obtained using the greedy algorithm
 \ref{alg:1} are presented under the column labeled {\textbf{2}}.

It is important to notice that for $K=2$ the Cordaro-Wagner codes are optimal and the greedy
approach indeed results in the same codes. Also for $K=3$ the solution obtained by our approach is
always optimal independently from the value of $d_{min}$. By increasing $K$ we diverge from the
optimal codes, however, for $K \leq 7$ the results are still surprisingly close to the optimal
codes. This observation, encourages us to use the same algorithm to design also rotationally
invariant codes as explained in the next sections. As a final note, it is important to notice that
the results in Table \ref{tab:codes1} provide us with a \emph{family} of rate compatible codes, as a good code
of length $N$ is obtained by adding a column to the generator matrix of a good code with length $N-1$.

On the other hand, if one is not interested in the rate compatibility constraint, one may run
multiple time the greedy algorithm and pick the optimum designed codes for any desired pair
$(K,d_{\mathrm{min}})$. The random selection of the  column in step 5 of the algorithm guarantees
that each run of the greedy algorithm yields different results. We show the best results for 100 runs in
Table \ref{tab:codes100}. The improvement compared to a single run is quite significant for $K=8,9$
and 10.

\begin{table}[htbp]
  \centering
 \caption{Code length for given $K$ and target minimum distance $d_{\mathrm{min}}$. Values are reported for best binary codes ({\textbf{B}}) and designed RC codes with greedy algorithm:
  Binary (\textbf{2}),   Binary and RI for BPSK (\textbf{2 RI2}),
  Binary and robust to NC detection for QPSK (\textbf{2 NC4}),
  over $\mathbb{Z}_4$ (\textbf{4}),over
  $\mathbb{Z}_4$ and RI for QPSK (\textbf{4 RI4}). All codes are obtained with a single run of the greedy algorithm.
  }     {\scriptsize
\setlength\tabcolsep{0.5mm}
    \begin{tabular}{|c|cccc|cc|c|ccc|cc|c|ccc|cc|}
    \hline
      & \multicolumn{1}{c|}{\textbf{B}} & \textbf{2} & \textbf{2 RI2} & \textbf{2 NC4} & \textbf{4} & \textbf{4 RI4} & \textbf{B} & \textbf{2} & \textbf{2 RI2} & \textbf{2 NC4} & \textbf{4} & \textbf{4 RI4} & \textbf{B} & \textbf{2} & \textbf{2 RI2} & \textbf{2 NC4} & \textbf{4} & \textbf{4 RI4} \\
    \hline
    $K$\textbackslash{}$d_{\mathrm{min}}$ & \multicolumn{6}{c|}{2} & \multicolumn{6}{c|}{4} & \multicolumn{6}{c|}{10} \\
    \hline
    2 & \textbf{3} & 3 & 4 & 4 & 4 & 4 & \textbf{6} & 6 & 8 & 8 & 6 & 8 & \textbf{15} & 15 & 20 & 20 & 16 & 20 \\
    3 & \textbf{4} & 4 & 4 & 4 & 6 & 4 & \textbf{7} & 7 & 8 & 8 & 8 & 8 & \textbf{18} & 18 & 20 & 20 & 20 & 20 \\
    4 & \textbf{5} & 5 & 6 & 6 & 6 & 6 & \textbf{8} & 9 & 9 & 8 & 10 & 8 & \textbf{20} & 20 & 22 & 22 & 22 & 22 \\
    5 & \textbf{6} & 6 & 6 & 7 & 8 & 8 & \textbf{10} & 10 & 10 & 11 & 10 & 12 & \textbf{21} & 22 & 22 & 23 & 24 & 24 \\
    6 & \textbf{7} & 7 & 8 & 8 & 8 & 8 & \textbf{11} & 11 & 12 & 12 & 14 & 14 & \textbf{23} & 23 & 24 & 24 & 26 & 24 \\
    7 & \textbf{8} & 8 & 8 & 8 & 10 & 10 & \textbf{12} & 12 & 14 & 14 & 16 & 14 & \textbf{24} & 25 & 26 & 26 & 28 & 28 \\
    8 & \textbf{9} & 9 & 10 & 10 & 10 & 10 & \textbf{13} & 14 & 14 & 14 & 16 & 16 & \textbf{26} & 27 & 28 & 28 & 30 & 30 \\
    9 & \textbf{10} & 10 & 10 & 11 & 12 & 12 & \textbf{14} & 15 & 16 & 16 & 18 & 16 & \textbf{27} & 28 & 28 & 30 & 32 & 32 \\
    10 & \textbf{11} & 11 & 12 & 12 & 12 & 12 & \textbf{15} & 16 & 18 & 16 & 18 & 18 & \textbf{28} & 30 & 30 & 31 & 34 & 34 \\
    11 & \textbf{12} & 12 & 12 & 12 & 14 & 14 & \textbf{16} & 17 & 18 & 18 & 20 & 18 & \textbf{30} & 31 & 32 & 32 & 38 & 36 \\
    12 & \textbf{13} & 13 & 14 & 14 & 14 & 14 & \textbf{18} & 18 & 18 & 19 & 20 & 20 & \textbf{31} & 33 & 34 & 34 & 36 & 36 \\
    13 & \textbf{14} & 14 & 14 & 15 & 16 & 16 & \textbf{19} & 19 & 20 & 20 & 22 & 22 & \textbf{32} & 35 & 36 & 36 & 38 & 38 \\
    14 & \textbf{15} & 15 & 16 & 16 & 16 & 16 & \textbf{20} & 20 & 21 & 20 & 24 & 24 & \textbf{34} & 35 & 36 & 36 & 42 & 40 \\
     \hline
    \end{tabular}
    \begin{tabular}{|c|c|ccc|cc|c|ccc|cc|c|ccc|cc|}
    \hline
      & \textbf{B} & \textbf{2} & \textbf{2 RI2} & \textbf{2 NC4} & \textbf{4} & \textbf{4 RI4} & \textbf{B} & \textbf{2} & \textbf{2 RI2} & \textbf{2 NC4} & \textbf{4} & \textbf{4 RI4} & \textbf{B} & \textbf{2} & \textbf{2 RI2} & \textbf{2 NC4} & \textbf{4} & \textbf{4 RI4} \\
    \hline
   $K$\textbackslash{}$d_{\mathrm{min}}$ & \multicolumn{6}{c|}{20} & \multicolumn{6}{c|}{30} & \multicolumn{6}{c|}{50} \\
    \hline
    2 & \textbf{30} & 30 & 40 & 40 & 30 & 40 & \textbf{45} & 45 & 60 & 60 & 46 & 60 & \textbf{75} & 75 & 100 & 100 & 76 & 100 \\
    3 & \textbf{35} & 35 & 40 & 40 & 36 & 40 & \textbf{53} & 53 & 60 & 60 & 54 & 60 & \textbf{88} & 88 & 100 & 100 & 88 & 100 \\
    4 & \textbf{38} & 39 & 40 & 40 & 40 & 40 & \textbf{57} & 57 & 62 & 62 & 60 & 62 & \textbf{95} & 95 & 102 & 102 & 96 & 102 \\
    5 & \textbf{40} & 41 & 42 & 43 & 42 & 44 & \textbf{59} & 61 & 62 & 63 & 62 & 64 & \textbf{99} & 100 & 102 & 103 & 104 & 104 \\
    6 & \textbf{42} & 43 & 44 & 44 & 46 & 46 & \textbf{60} & 63 & 66 & 64 & 68 & 64 & \textbf{101} & 103 & 104 & 104 & 110 & 106 \\
    7 & \textbf{43} & 45 & 46 & 47 & 48 & 50 & \textbf{62} & 66 & 66 & 64 & 70 & 70 & \textbf{102} & 107 & 106 & 108 & 112 & 110 \\
    8 & \textbf{45} & 49 & 50 & 48 & 52 & 52 & \textbf{65} & 69 & 71 & 68 & 76 & 72 & \textbf{105} & 110 & 110 & 110 & 118 & 112 \\
    9 & \textbf{47} & 50 & 50 & 52 & 54 & 54 & \textbf{67} & 71 & 72 & 72 & 74 & 72 & \textbf{107} & 112 & 114 & 115 & 118 & 116 \\
    10 & \textbf{48} & 52 & 54 & 54 & 56 & 56 & \textbf{68} & 74 & 76 & 76 & 80 & 78 & \textbf{109} & 116 & 118 & 118 & 122 & 120 \\
    11 & \textbf{50} & 55 & 54 & 56 & 62 & 56 & \textbf{70} & 77 & 78 & 78 & 84 & 80 & \textbf{113} & 120 & 120 & 120 & 128 & 126 \\
    12 & \textbf{52} & 56 & 56 & 58 & 64 & 60 & \textbf{72} & 78 & 80 & 80 & 86 & 84 & \textbf{116} & 122 & 124 & 124 & 130 & 128 \\
    13 & \textbf{53} & 58 & 60 & 60 & 62 & 64 & \textbf{74} & 80 & 82 & 84 & 88 & 86 & \textbf{118} & 124 & 126 & 127 & 134 & 130 \\
    14 & \textbf{54} & 60 & 60 & 60 & 62 & 64 & \textbf{76} & 82 & 84 & 84 & 88 & 88 & \textbf{119} & 128 & 128 & 128 & 134 & 130 \\
    \hline
    \end{tabular}%
    }%
  \label{tab:codes1}%
\end{table}%

\begin{table}[htbp]
  \centering
  \caption{Code length for given $K$ and target minimum distance $d_{\mathrm{min}}$. Values are reported for best binary codes ({\textbf{B}}) and several designed RC codes:
  Binary (\textbf{2}),   Binary and RI for BPSK (\textbf{2 RI2}),
  Binary and robust to NC detection for QPSK (\textbf{2 NC4}),
  over $\mathbb{Z}_4$ (\textbf{4}),over
  $\mathbb{Z}_4$ and RI  for QPSK (\textbf{4 RI4}). The values of best codes with 100 runs of the greedy algorithm are reported.}
     {\scriptsize
\setlength\tabcolsep{0.5mm}
    \begin{tabular}{|r|c|ccc|cc|c|ccc|cc|c|ccc|cc|}
     \hline
      & \textbf{B} & \textbf{2} & \textbf{2 RI2} & \textbf{2 NC4} & \textbf{4} & \textbf{4 RI4} & \textbf{B} & \textbf{2} & \textbf{2 RI2} & \textbf{2 RI4} & \textbf{4} & \textbf{4 RI4} & \textbf{Best} & \textbf{2} & \textbf{2 RI2} & \textbf{2 RI4} & \textbf{4} & \textbf{4 RI4} \\
     \hline
    \multicolumn{1}{|l|}{$K$\textbackslash{}$d_{min}$} & \multicolumn{6}{c|}{2} & \multicolumn{6}{c|}{4} & \multicolumn{6}{c|}{10} \\
    \hline
    2 & \textbf{3} & 3 & 4 & 4 & 4 & 4 & \textbf{6} & 6 & 8 & 8 & 6 & 8 & \textbf{15} & 15 & 20 & 20 & 16 & 20 \\
    3 & \textbf{4} & 4 & 4 & 4 & 4 & 4 & \textbf{7} & 7 & 8 & 8 & 8 & 8 & \textbf{18} & 18 & 20 & 20 & 18 & 20 \\
    4 & \textbf{5} & 5 & 6 & 6 & 6 & 6 & \textbf{8} & 9 & 8 & 8 & 10 & 8 & \textbf{20} & 20 & 22 & 22 & 22 & 22 \\
    5 & \textbf{6} & 6 & 6 & 7 & 6 & 8 & \textbf{10} & 10 & 10 & 11 & 10 & 12 & \textbf{21} & 21 & 22 & 23 & 22 & 24 \\
    6 & \textbf{7} & 7 & 8 & 8 & 8 & 8 & \textbf{11} & 11 & 12 & 12 & 12 & 12 & \textbf{23} & 23 & 24 & 24 & 24 & 24 \\
    7 & \textbf{8} & 8 & 8 & 8 & 8 & 8 & \textbf{12} & 12 & 13 & 13 & 12 & 14 & \textbf{24} & 25 & 26 & 26 & 26 & 26 \\
    8 & \textbf{9} & 9 & 10 & 10 & 10 & 10 & \textbf{13} & 14 & 14 & 14 & 14 & 16 & \textbf{26} & 26 & 28 & 28 & 28 & 28 \\
    9 & \textbf{10} & 10 & 10 & 11 & 10 & 12 & \textbf{14} & 15 & 15 & 15 & 14 & 16 & \textbf{27} & 28 & 28 & 28 & 30 & 28 \\
    10 & \textbf{11} & 11 & 12 & 12 & 12 & 12 & \textbf{15} & 16 & 16 & 16 & 18 & 16 & \textbf{28} & 29 & 30 & 30 & 32 & 30 \\
     \hline
    \end{tabular}%

    \begin{tabular}{|c|c|ccc|cc|c|ccc|cc|c|ccc|cc|}
     \hline
     & \textbf{B} & \textbf{2} & \textbf{2 RI2} & \textbf{2 NC4} & \textbf{4} & \textbf{4 RI4} & \textbf{B} & \textbf{2} & \textbf{2 RI2} & \textbf{2 RI4} & \textbf{4} & \textbf{4 RI4} & \textbf{B} & \textbf{2} & \textbf{2 RI2} & \textbf{2 RI4} & \textbf{4} & \textbf{4 RI4} \\
    \hline
    $K$\textbackslash{}$d_{min}$ & \multicolumn{6}{c|}{20} & \multicolumn{6}{c|}{30} & \multicolumn{6}{c|}{50} \\
    \hline
    2 & \textbf{30} & 30 & 40 & 40 & 30 & 40 & \textbf{45} & 45 & 60 & 60 & 46 & 60 & \textbf{75} & 75 & 100 & 100 & 76 & 100 \\
    3 & \textbf{35} & 35 & 40 & 40 & 36 & 40 & \textbf{53} & 53 & 60 & 60 & 54 & 60 & \textbf{88} & 88 & 100 & 100 & 88 & 100 \\
    4 & \textbf{38} & 38 & 40 & 40 & 38 & 40 & \textbf{57} & 57 & 62 & 62 & 58 & 62 & \textbf{95} & 95 & 102 & 102 & 96 & 102 \\
    5 & \textbf{40} & 40 & 42 & 43 & 42 & 44 & \textbf{59} & 59 & 62 & 62 & 60 & 62 & \textbf{99} & 99 & 102 & 103 & 100 & 104 \\
    6 & \textbf{42} & 42 & 44 & 44 & 44 & 44 & \textbf{60} & 62 & 62 & 62 & 64 & 62 & \textbf{101} & 102 & 104 & 104 & 104 & 104 \\
    7 & \textbf{43} & 45 & 44 & 46 & 46 & 46 & \textbf{62} & 65 & 64 & 64 & 68 & 62 & \textbf{102} & 105 & 106 & 106 & 108 & 106 \\
    8 & \textbf{45} & 47 & 48 & 48 & 50 & 48 & \textbf{65} & 68 & 68 & 68 & 70 & 68 & \textbf{105} & 109 & 108 & 108 & 114 & 110 \\
    9 & \textbf{47} & 49 & 50 & 51 & 52 & 52 & \textbf{67} & 70 & 72 & 72 & 74 & 72 & \textbf{107} & 112 & 114 & 114 & 116 & 114 \\
    10 & \textbf{48} & 51 & 52 & 52 & 54 & 52 & \textbf{68} & 73 & 74 & 74 & 78 & 76 & \textbf{109} & 115 & 116 & 116 & 120 & 118 \\
     \hline
    \end{tabular}}%
  \label{tab:codes100}%
\end{table}%

\subsection{Construction of rate-compatible linear  codes over $\mathbb{Z}_M$}\label{sec:algorithm2} The
previous greedy algorithm for construction of good rate compatible linear binary codes
can be generalized to construct linear codes over $\mathbb{Z}_M$. In this section
we consider codes over $\mathbb{Z}_4$ while extension to larger values of $M$ is straightforward.

Any $(K,N)$ linear code $\mathcal{C}$ over $\mathbb{Z}_4$ (sub-module) can be generated with  a
generator matrix of the form:
\[
\mathbf{G}=
\begin{pmatrix}
  A \\
  2B \\
\end{pmatrix},
\]
where $A$ is a $(k_1\times N/2)$ matrix  with elements in $\mathbb{Z}_4$ and $B$ is a $(k_2\times N/2)$
matrix  with elements in $\mathbb{Z}_2$. The code, which has dimension $K=2k_1+ k_2$, is generated
by multiplying $\mathbf{G}$ by a vector $\mathbf{u}^T=(\mathbf{u}_1,\mathbf{u}_2)$ with the first
$k_1$ components in $\mathbb{Z}_4$ and the last $k_2$ components in $\mathbb{Z}_2$. Multiple choice
of the pair $k_1,k_2$ for the same $K$ originate different code types.

If the considered distance between group elements is such that it can be computed by applying a proper
weight function $w_4$ to their $\mathbb{Z}_4$ sum
\[
d(a,b)= w_4(a+ b),
\]
linearity of code ($\forall \mathbf{a},\mathbf{b} \in \mathcal{C}\rightarrow \mathbf{a}+
\mathbf{b}\in \mathcal{C}$) implies that distance spectrum from any codeword coincides with the
weight spectrum of the code. Considering these generalizations, step 4 of the greedy algorithm
\ref{alg:1} should be substituted to
          \begin{itemize}
          \item Construct the list of  input words $\mathcal{U}\in
              \mathbb{Z}^{k_1}_4\times\mathbb{Z}^{k_2}_2$ generating the minimum weight
              codewords (nearest neighbors):
          \[   \mathcal{U}=\left\{\mathbf{u}=\arg \min \big( w_4(\mathbf{u^TG}^{(n)}) \big) \right\}\]
          \end{itemize}

and step 5 with
          \begin{itemize}
          \item Find a new column vector $\mathbf{g}^{(n+1)}$, to be appended to
              $\mathbf{G}^{(n)}$,  that reduces as much as possible the number of nearest
              neighbors:
         \begin{equation}
          \mathbf{g}^{(n+1)} \in \arg\min_{\mathbf{g}}\sum_{\mathbf{u}\in \mathcal{U}} \mathbb{I}(0= w_4(\mathbf{g^T\cdot u}))\;\;
          \label{eq:opt1},
         \end{equation}
         where $\mathbb{I}()$ is the indicator function, returning 1 if its argument is true.

           \end{itemize}
With the natural $M$-PSK mapping $m\leftrightarrow \frac{1}{\sqrt{2}}\mathrm{e}^{j2\pi m/M}$ the
induced weight correspondent to the Euclidean distance is $w_4(0,1,2,3)=(0,1,2,1)$ so that
\cref{eq:opt1} is not equivalent to \cref{eq:opt}. When multiple  columns exist satisfying
\cref{eq:opt1} one additional requirement may be that of minimizing the number of scalar products with weight
1.

The results for constructed codes over $\mathbb{Z}_4$ are presented in
\cref{tab:codes1,tab:codes100} under the column label {\textbf{4}}.
For each value of $K$ we report the results for the code type with the largest value of $k_1$, i.e. $k_1=\lfloor K/2 \rfloor$.
Results for differents type are not reported as they do not provide significant differences.

 Notice that the length in bits $N$ of codes over $\mathbb{Z}_4$ is always an even number. For small values of $K$ the codes
$\mathbb{Z}_4$ have the same length as their binary counterparts (with exception of 1 bit due to
the even length constraint). On the other hand, for large $K$ a small increase can be seen.

\subsection{Construction of rotationally invariant codes}\label{sec:ricode} A codebook
$\mathcal{S}$ for $M$-PSK constellation is rotationally invariant when all the rotated versions
 of any codeword belong to the codebook.

For $M$-PSK constellation and linear codes over $\mathbb{Z}_M$ generating group, this condition
is equivalent to impose that the all-one  codeword  (denoted by  $\overline{1}$) belong to the codebook
$\mathcal{C}$. This property can be enforced constraining the first row of the generating matrix
$\mathbf{G}$ to $\overline{1}$. This additional constraint in turn can be easily incorporated into
the greedy Algorithm \ref{alg:1} to generate families of good rate compatible and rotational
invariant codes.

The results for rotationally invariant codes over $\mathbb{Z}_2$ and $\mathbb{Z}_4$ are presented
in \cref{tab:codes1,tab:codes100} with column labels {\textbf{2 RI2}} and {{\textbf{4 RI4}}
respectively. In principle the length of rotational invariant codes for a given minimum distance
and information bit must be larger or equal to those codes without any constraints. In our tables
in few cases when $K \leq 7$ the rotational invariant codes have smaller length compared to
optimized codes. This is due to two facts. First, the greedy algorithm does not always provides the
optimal solution and second, the search space is smaller for rotationally invariant codes (for any
given $K$) and therefore the greedy algorithm performs slightly better.

In Figure \ref{fig:distvsn} we report the minimum distance growth as a function of $K$ and $N$ of the
generated codes for BPSK and QPSK constellations with 100 runs of the greedy algorithm. The main observation is that as $K$ increases,
both the optimized codes and rotationally invariant codes show the same minimum distance growth as
a function of code length.

Generator matrices of a good rate compatible family corresponding to the row $K=8$ of \cref{tab:codes1} are reported in the appendix.
          \begin{figure}[h]
          \subfigure[Optimized code over $\mathbb{Z}_2$ ({\textbf{2}})]{\includegraphics[width=0.5\textwidth]{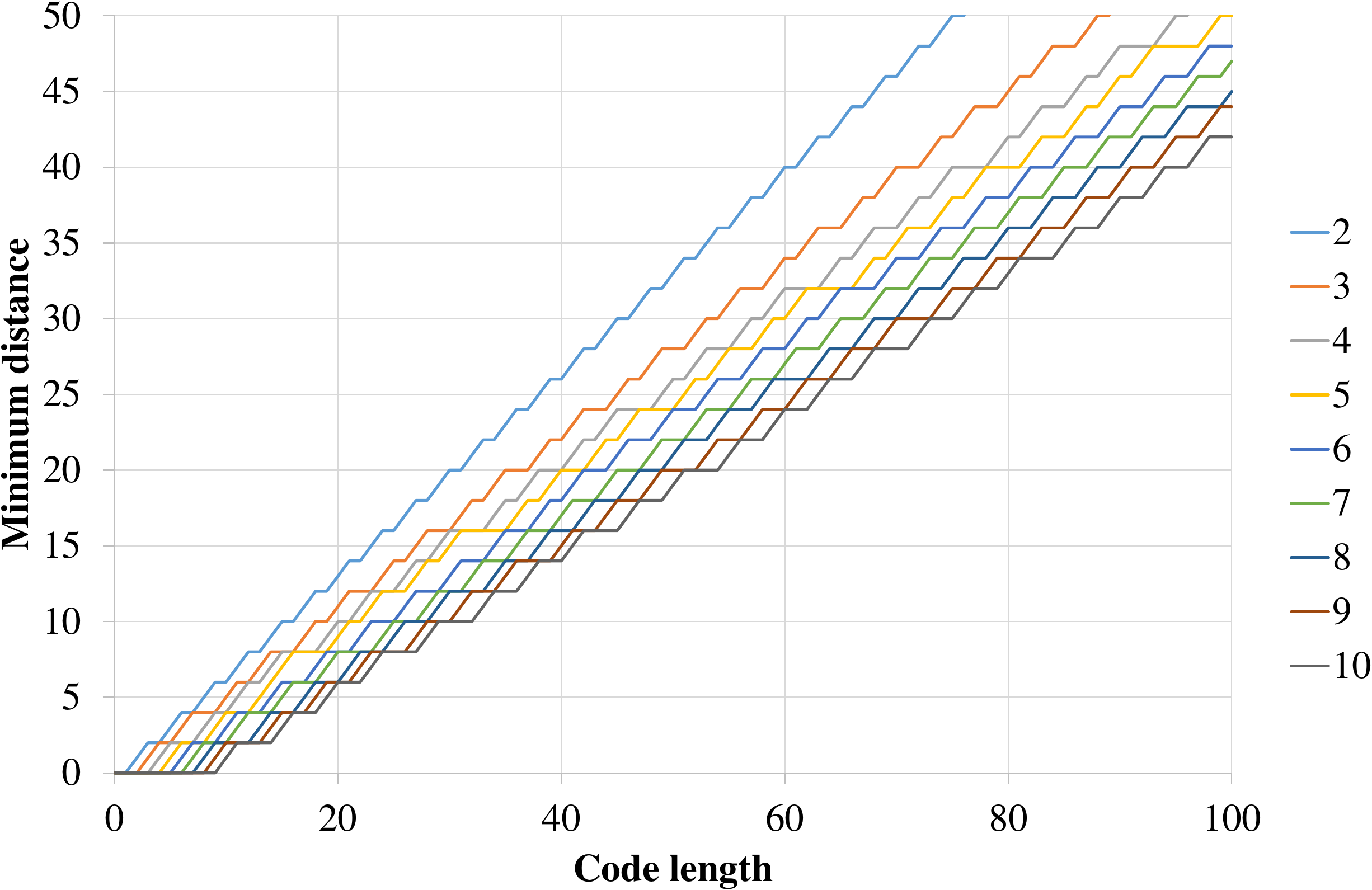}\label{fig:a}}
          \subfigure[Optimized code over $\mathbb{Z}_2$, RI  for BPSK constellation ({\textbf{2RI2)}}]{\includegraphics[width=0.5\textwidth]{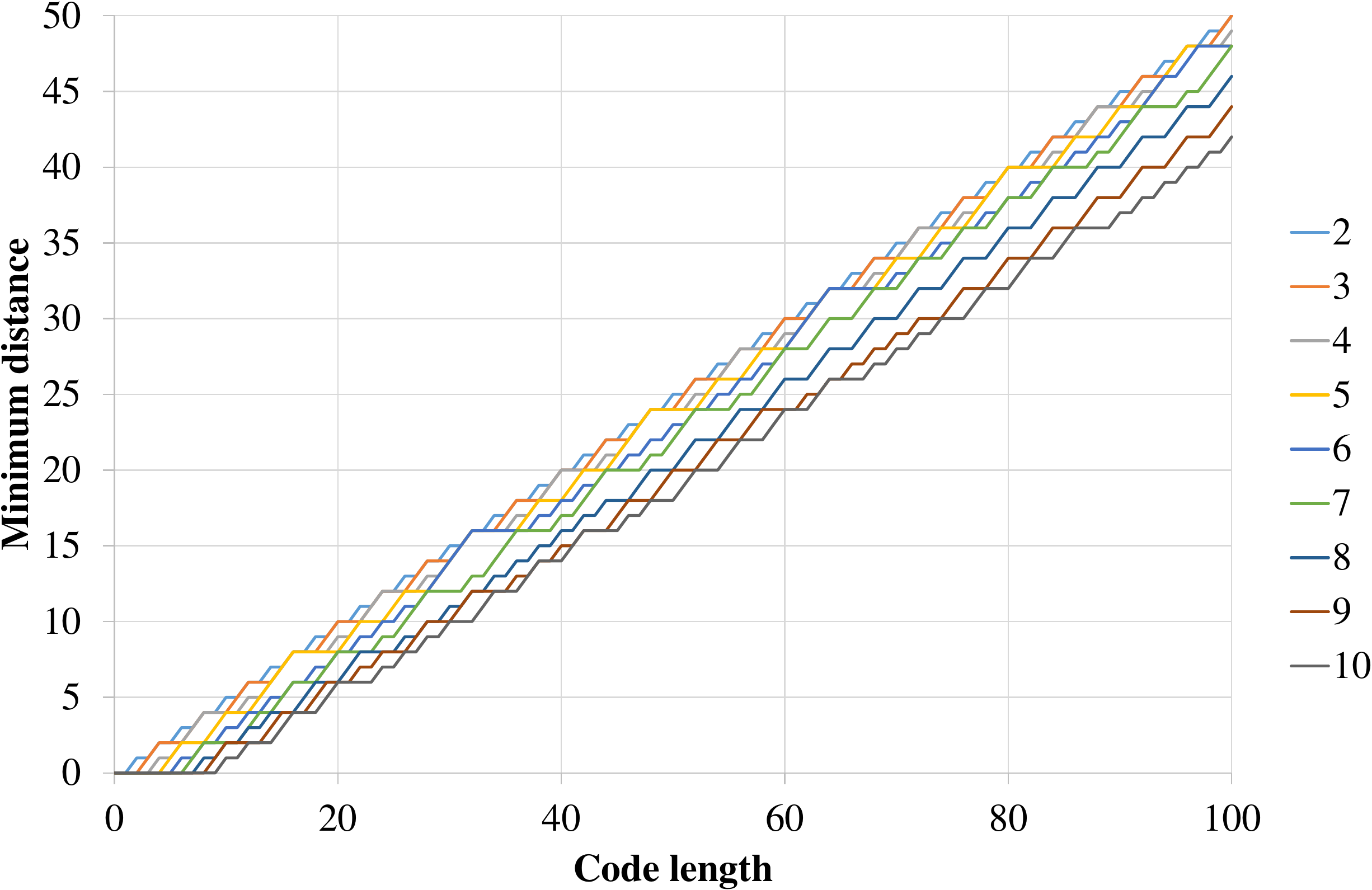}\label{fig:b}}\\
          \subfigure[Optimized code over $\mathbb{Z}_4$ ({\textbf{4}})]{\includegraphics[width=0.5\textwidth]{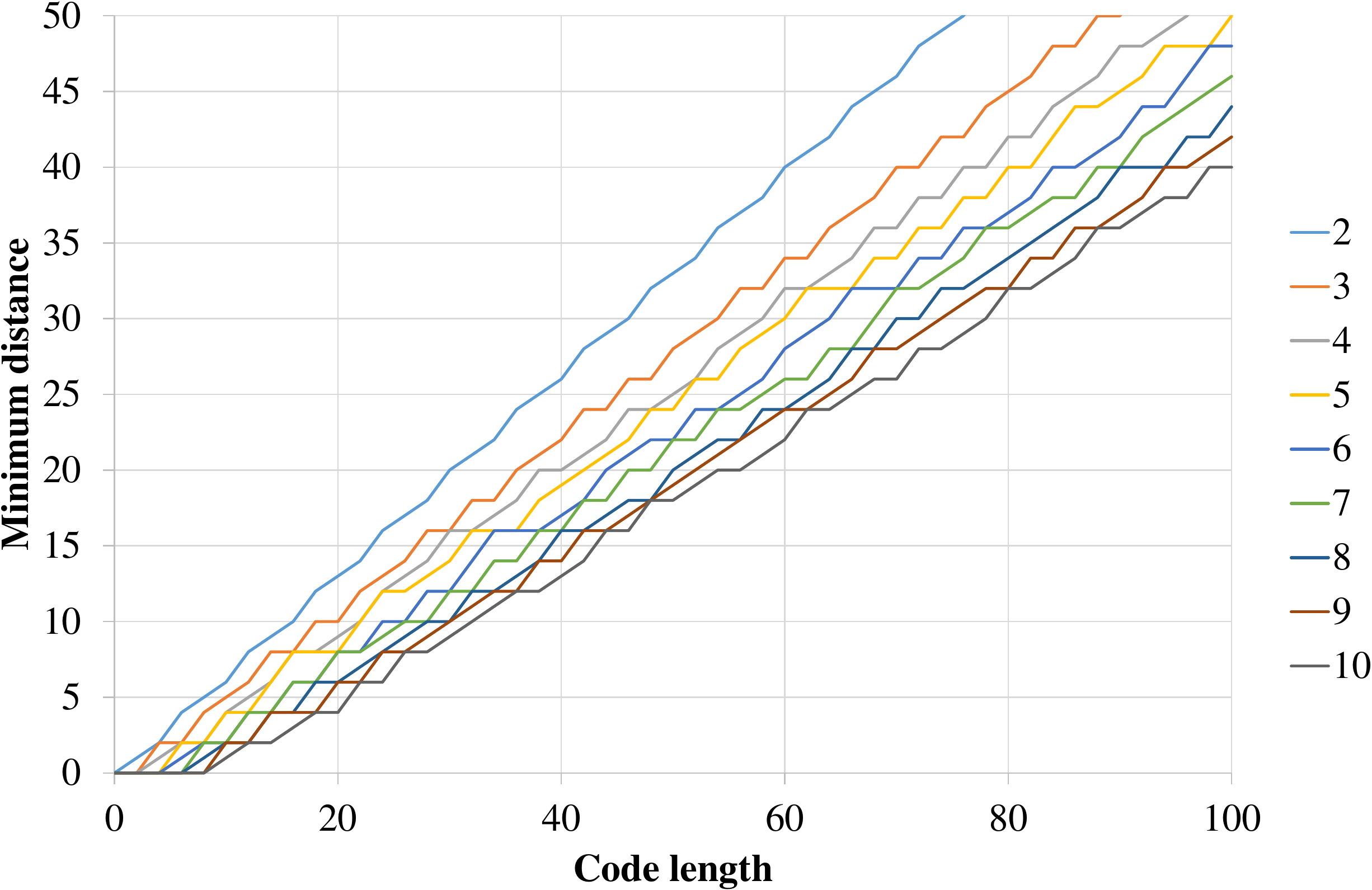}\label{fig:c}}
          \subfigure[Optimized code over $\mathbb{Z}_4$, RI for QPSK constellation ({\textbf{4RI4)}}]{\includegraphics[width=0.5\textwidth]{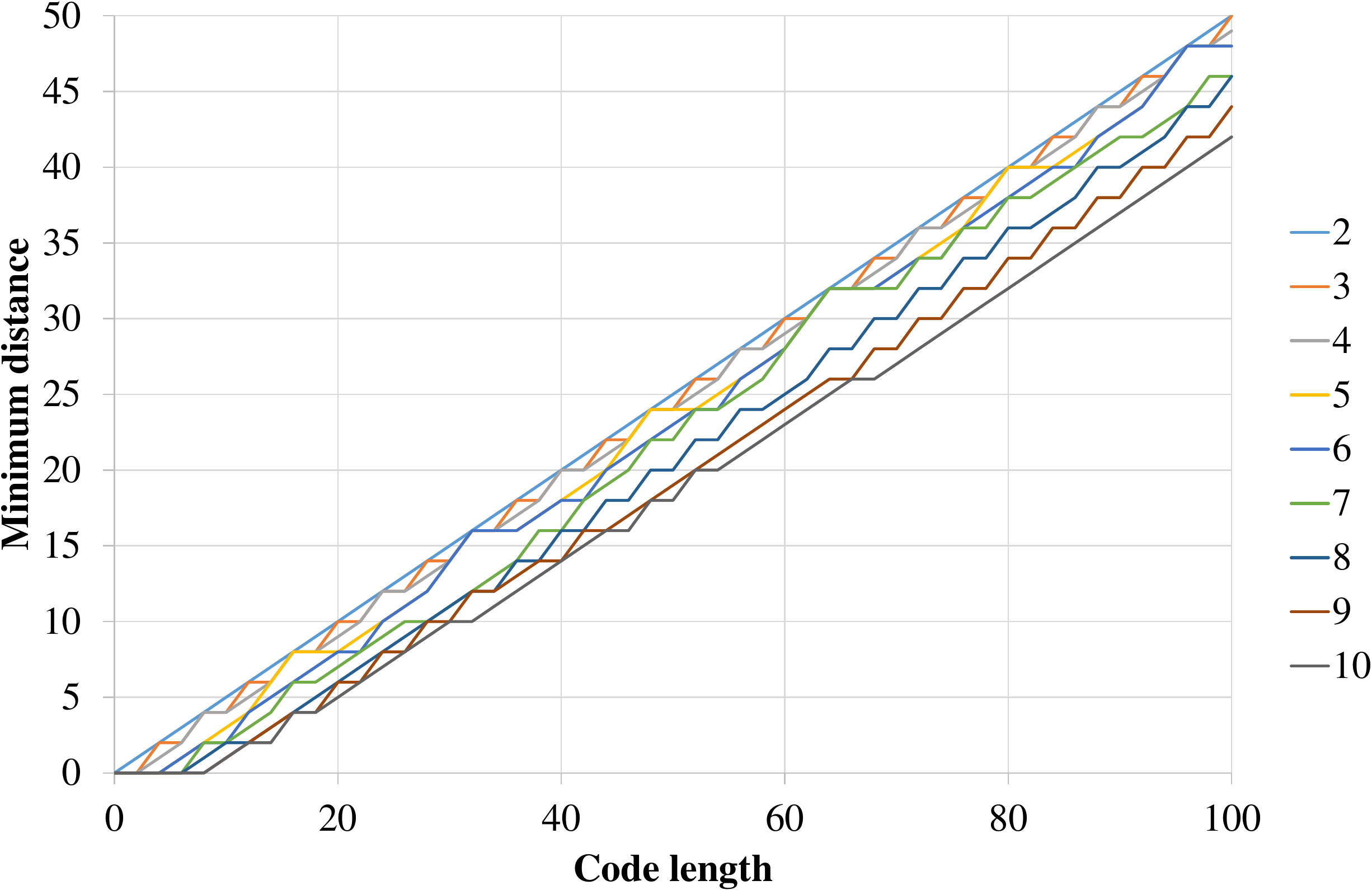}\label{fig:d}}
          \caption{Minimum distance growth as a function of code length for best codes obtained by greedy algorithm (100 runs).
          Several values of $K$ are reported.}
          \label{fig:distvsn}
          \end{figure}

\section{From rotationally invariant codes to codes for non-coherent detection}\label{sec:upperbound}
In this section we describe how to construct codes robust to non-coherent detection starting from
RI codes. First we present a simple upper bound on the FER of the RI codes which motivates our construction of the codes for non-coherent detection
\subsection{Union bound for non-coherent detection}
Simple upper bounds (UBs) to frame error probability of codes, $P_E$,  can be obtained by applying the union
bound based on the pairwise error probability (PEP) $P(\mathbf{s\rightarrow s'})$:
\begin{equation}\label{eq:UB}
 P_E\leq \frac{1}{2^K}\sum_{\mathbf{s}} \sum_{\mathbf{s'\neq s}}P(\mathbf{s\rightarrow s'}).
\end{equation}

For coherent detection and AWGN channel PEP is upper bounded by
\[ P(\mathbf{s\rightarrow
s'}) \leq
\exp\left(-\frac{||\mathbf{s-s'}||^2}{4N_0}\right),
\]
where $\mathbf{s,s'}$ are the constellation sequences in the codebook.

When codebook is constructed by applying a proper mapping $\mathbf{s(c)}$ to codewords $\mathbf{c}$ of linear code
$\mathcal{C}$ over a \emph{generating group} of the considered constellation the PEP can be
simplified as
\[
P(\mathbf{s(c)\rightarrow s(c')})=P(\mathbf{s(0)\rightarrow s(c'- c)}).
\]
In this case outer sum in (\ref{eq:UB}) can be eliminated (geometrical uniform codes) and UB can be written as:
          \begin{equation}
          P_E^c(\mathcal{C})\leq \sum_{\mathbf{c\in \mathcal{C}\neq 0}} P(\mathbf{s(0) \rightarrow s(c)})\label{eq:ubc}.
          \end{equation}
When using non-coherent detection on $M$-PSK, it can be demonstrated  (see for example
\cite{Colavolpe1}) that the PEP is asymptotically upper bounded (for large code lengths) by
          \begin{equation} \label{eq:3}
          P(\mathbf{s}\rightarrow \mathbf{s'}) \tilde{\leq} \exp\left(-\frac{d^2_{\mathrm{eq}}(\mathbf{s,s'})}{4N_0}\right),
          \end{equation}
where
\[
d^2_{\mathrm{eq}}(\mathbf{s,s'})=\min_{i=0,\ldots,M-1}||\mathbf{s}-e^{j \frac{2\pi i}{M}}\mathbf{s'}||^2,
\]
and $\tilde{\leq}$ indicates the asymptotic inequality as $N\rightarrow \infty$. Equation (\ref{eq:3}) in turn can be further upper bounded as follows
\[
P(\mathbf{s}\rightarrow \mathbf{s'})\tilde{\leq} \sum_{i=0}^{M-1} \exp\left(-\frac{||\mathbf{s}-e^{j 2\pi i/M}\mathbf{s'}||^2}{4N_0}\right).
 \]
The union bound for non-coherent detection takes then the following approximate form:
\begin{eqnarray}\label{eq:upperbound}
P_E^{nc}&\tilde{\leq}& \frac{1}{2^{K}}\sum_{\mathbf{s}}\sum_{\mathbf{s'\neq s}}\sum_{i=0}^{M-1} \exp\left(-\frac{||\mathbf{s}-e^{j 2\pi i/M}\mathbf{s'}||^2}{4N_0}\right).
\end{eqnarray}

Now consider a $(N,K-m)$ code $\mathcal{C}$ constructed from a linear $(N,K)$ RI code
$\mathcal{C}''$ over $\mathbb{Z}_M$ by eliminating the all-one row of its generating matrix. Its
union bound reads:
          \begin{eqnarray}
          P_E^{nc}(\mathcal{C})&\tilde{\leq}&
          \frac{1}{2^{K-m}}\sum_{\mathbf{c}}\sum_{\mathbf{c'\neq c}}\sum_{i=0}^{M-1} \exp\left(-\frac{||\mathbf{s(c)}-e^{j 2\pi i/M}\mathbf{s(c')}||^2}{4N_0}\right)\nonumber\\
          &=&
          \frac{1}{2^{K-m}}\sum_{\mathbf{c}}\sum_{\mathbf{c'\neq c}}\sum_{i=0}^{M-1} \exp\left(-\frac{||\mathbf{s(c)}-\mathbf{s(c'+ i)}||^2}{4N_0}\right).\label{eq:xx}
          \end{eqnarray}
The second identity stems from the fact that rotating the constellation sequence by  $2\pi i/M$
corresponds to adding  the all-$i$ sequence ($\mathbf{i}$) to the codeword. The two inner sums in
(\ref{eq:xx}) enumerate  all the codewords of $\mathcal{C}''$ not equal to a rotated version of
$\mathbf{s(c)}$, so that:
\begin{eqnarray}
P_E^{nc}(\mathcal{C})&\tilde{\leq}&
\frac{1}{2^{K-m}}\sum_{\mathbf{c}}\sum_{\mathbf{c''\in \mathcal{C}''},\mathbf{c''} \neq \mathbf{c + i}}
\exp\left(-\frac{||\mathbf{s(c)-s(c'')}||^2}{4N_0}\right).\label{eq:uu}
\end{eqnarray}
 Now we can exploit the geometrical uniformity of $\mathcal{C}$ to simplify the UB (\ref{eq:uu}) removing
the outer sum:
\begin{eqnarray}
P_E^{nc}(\mathcal{C})&\tilde{\leq}&
\sum_{\mathbf{c''\in \mathcal{C}''} \neq \mathbf{i}}
\exp\left(-\frac{||\mathbf{s(0)-s(c'')}||^2}{4N_0}\right)\leq P_E^{c}(\mathcal{C''})\label{eq:ubnc}
\end{eqnarray}

An approximation of the upper bound to FER with non-coherent detection for the $(N,K-m)$ code
$\mathcal{C}$ is then upper bounded  by the upper bound to FER with coherent detection of the
correspondent RI code $\mathcal{C}''$. The difference being that all $M$ rotated sequences
$\mathbf{i}$ are excluded from the sum (\ref{eq:ubnc}) instead of just the all zero sequence as in
(\ref{eq:ubc}).

The generating matrix of a \emph{good} $(K-m,N)$ code for non-coherent detection of $2^m$-PSK can
then be obtained starting  with a \emph{good} $K\times N$ generating  matrix of a rotationally invariant
code constructed as described in section~\ref{sec:ricode} by eliminating the first all-one row. The
new codebook is such that the distance from any codeword to any other codeword \emph{and all its
rotated versions} is large.

\subsection{Codes on $\mathbb{Z}_2$ for non-coherent detection of QPSK constellations}
Even though linear codes built over $\mathbb{Z}_M$ can be represented in binary, they are not
binary linear codes. Focusing on QPSK constellation, our previous construction results in a linear
code over $\mathbb{Z}_4$. If a linear \emph{binary} code is desired one can use the alternative
construction  presented in \cite{ASMSerror}. In this case one construct a \emph{binary} $K \times
N$ generator matrix where the two first rows are fixed to be $\overline{10}$ and $\overline{01}$
codewords using the above mentioned greedy approach. This forces the binary codebook to have
$\overline{1}$, $\overline{10}$ and $\overline{01}$ as codewords. However, since $\mathbb{Z}_4$ and
$\mathbb{Z}^2_2$ are not isomorphic, no direct mapping can be found between the rotations of QPSK
constellation and the codebook algebra. Therefore, contrary to what has been reported previously in
\cite{ASMSerror}, these codes are not rotationally invariant. Nevertheless, code suitable for
non-coherent detection can still be obtained by this construction after eliminating the first two
rows. Our simulations indicate that such codes perform quite well with the non-coherent detector.
The results based on this construction are reported in Tables \ref{tab:codes1} and
\ref{tab:codes100} under the column name {\textbf{2 NC4}}.

\section{Example of PLH code construction for BPSK and QPSK constellations}\label{sec:simul}
In this section we report an example of PLH code construction for $K=8$  and target FER=$10^{-8}$
for SNR=0, 5 and 10 dB.

The first step is that of establishing the required minimum distance of codes for achieving the
desired FER at the target SNR. In order to do so we use the following conservative upper bound to
FER
\begin{equation}
FER\leq (2^{K}-1)\frac{1}{2}\mathrm{erfc}\left(\sqrt{Ad_{\mathrm{min}}\frac{E_s }{N_0}}\right).\label{eq:ubdmin}
 \end{equation}
with $A=1$ for BPSK and $A=1/2$ for QPSK. This simple bound  provides a conservative value for the minimum distance using
BPSK and QPSK. The required values are reported in 2nd and 5th column of \cref{tab:tabd}.

The required generating matrices are then selected depending on the type of detection. When using
coherent detection (CD) one has to pick the generating matrix of a $(N,8)$ code with minimal length $N$ achieving the required
minimum distance. When using non-coherent detection (NCD)  one has to pick  the generating matrix of a  $(N,8+m)$ RI code over $\mathbb{Z}_{2^m}$
with minimal length and remove the first row. The required lengths (in symbols) using the rate-compatible
code families constructed as described in section \ref{sec:construct} are reported in
\cref{tab:tabd}. One can see that, as expected,  the additional requirement of robustness to non-coherent
detection has a marginal impact on the required code lengths. One can also observe that using QPSK
gives a small advantage in terms of required PLH length in symbols.

\begin{table}[htbp]
  \centering
  \caption{Required code lengths (symbols) for PLH achieving FER=$10^{-8}$ at 0, 5 and 10 dB. \textbf{CD}: Coherent Detection. \textbf{NCD}: Non-Coherent Detection.}
    \begin{tabular}{|r|rrr|rrr|}
   \hline
      & \multicolumn{3}{c|}{\textbf{BPSK}} & \multicolumn{3}{c|}{\textbf{QPSK}} \\
   \hline
    \multicolumn{1}{|l|}{\textbf{SNR}} & \multicolumn{1}{l}{$d_{\mathrm{min}}$} & \multicolumn{1}{l}{\textbf{CD}} & \multicolumn{1}{l|}{\textbf{NCD}}
    & \multicolumn{1}{l}{$d_{\mathrm{min}}$} & \multicolumn{1}{l}{\textbf{CD}} & \multicolumn{1}{l|}{\textbf{NCD}} \\
   \hline
    \textbf{0} & 22 & 51 & 54 & 43 & 50 & 52  \\
    \textbf{5} & 7  & 21 & 22 & 14 & 19 & 20   \\
    \textbf{10}& 3  & 13 & 14 & 5 & 9  & 10   \\
    \hline
    \end{tabular}%
  \label{tab:tabd}%
\end{table}%

We simulated the FER performance of codes constructed in this way when using coherent ML detection and noncoherent ML detection.
In Figures \ref{fig:BPSKNRI} to \ref{fig:QPSKRI} we report the performances of designed codes  and comparison with bounds
(\ref{eq:ubc}) or (\ref{eq:ubnc})  (dashed lines)  or the simpler bound (\ref{eq:ubdmin}) (solid lines).
\begin{figure}[h]\center
  \includegraphics[width=.85\linewidth]{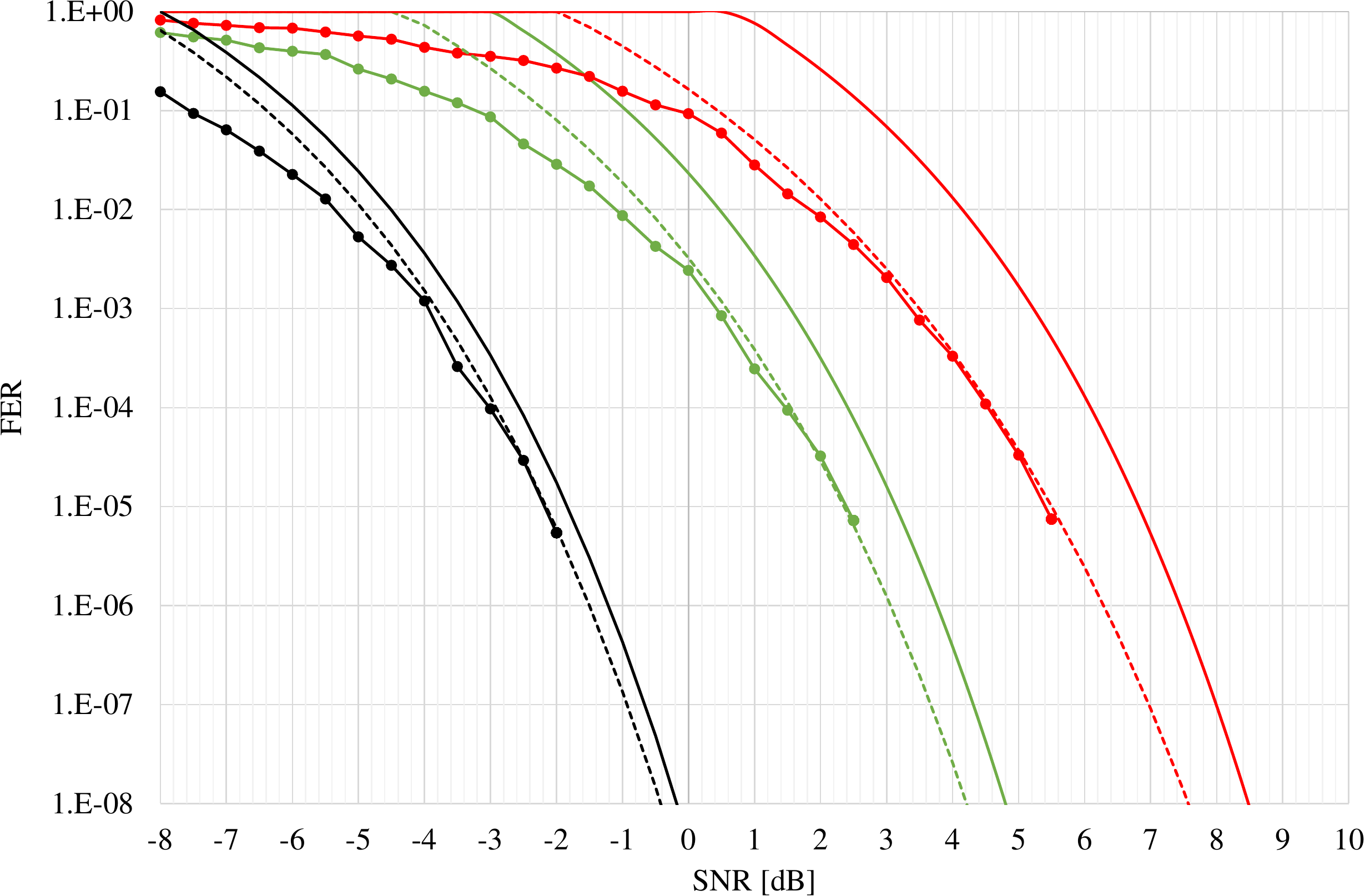}\\
  \caption{Comparison of simulation results and bounds for codes designed for BPSK with coherent detection \textbf{(2)}.
  Target SNR at FER=$10^{-8}$ is 0, 5 and 10 dB, corresponding to required $d_{\mathrm{min}}$ of 22,7, and 3.}\label{fig:BPSKNRI}
\end{figure}
\begin{figure}[h]\center
  \includegraphics[width=.85\linewidth]{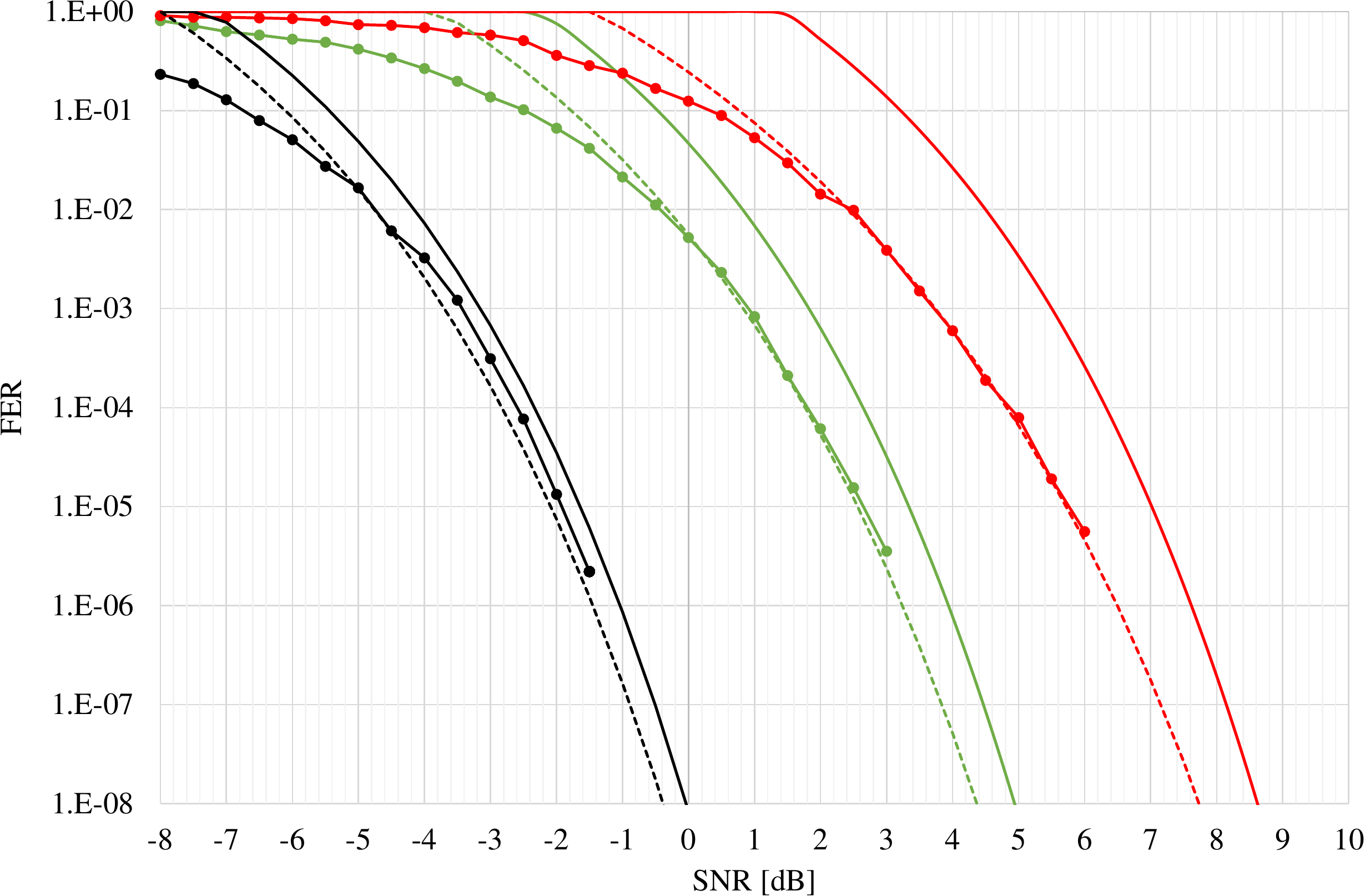}\\
  \caption{Comparison of simulation results and bounds for codes designed for BPSK with non-coherent detection ({\textbf{2RI2}}).
  Target SNR at FER=$10^{-8}$ is 0, 5 and 10 dB, corresponding to required $d_{\mathrm{min}}$ of 22,7, and 3 respectively.}\label{fig:BPSKRI}
\end{figure}
\begin{figure}[h]\center
  \includegraphics[width=.85\linewidth]{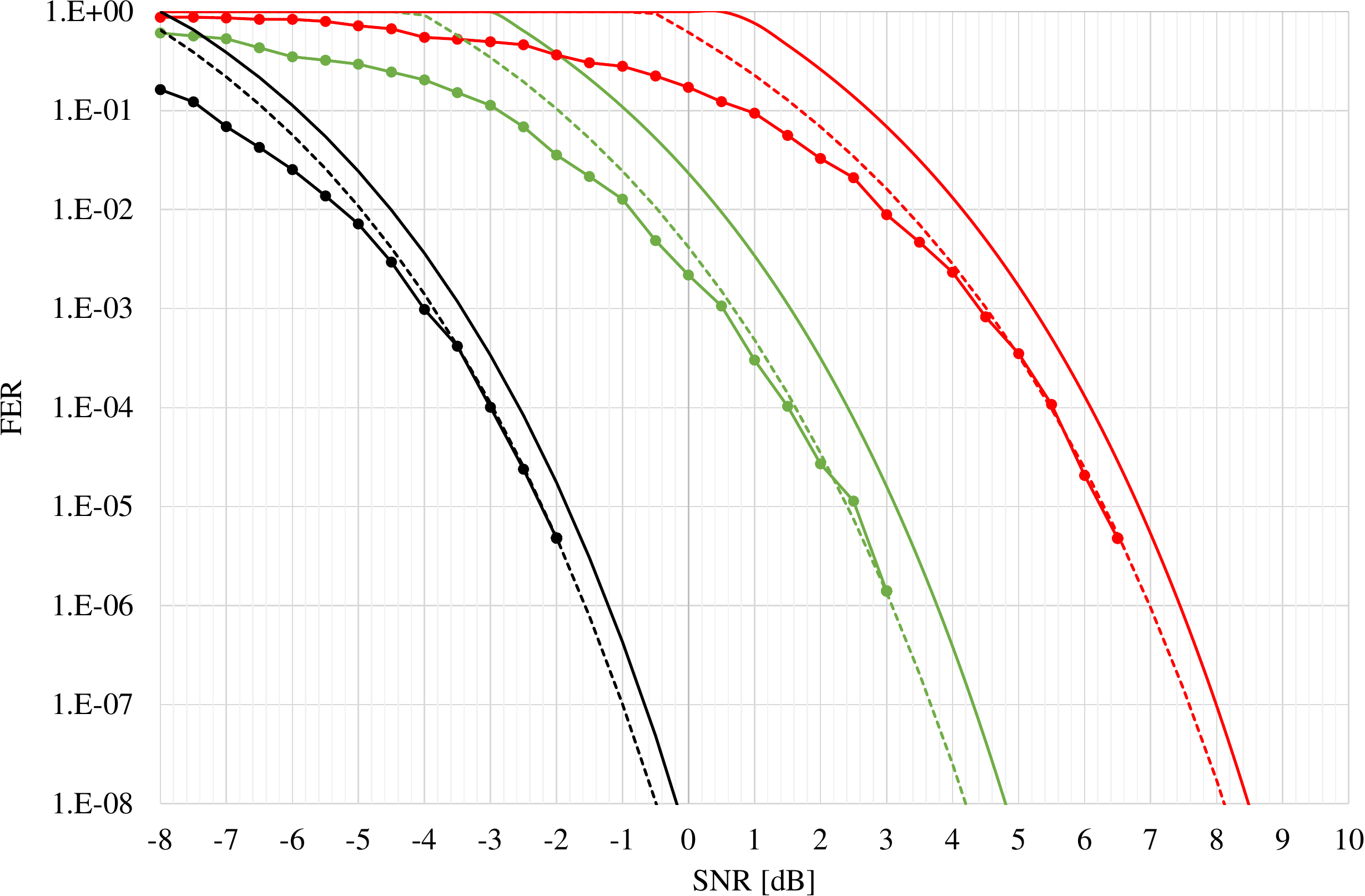}\\
  \caption{Comparison of simulation results and bounds for codes designed for QPSK with coherent detection ({\textbf{4}}).
  Target SNR at FER=$10^{-8}$ is 0, 5 and 10 dB, corresponding to required $d_{\mathrm{min}}$ of 43,14, and 5.}\label{fig:QPSKNRI}
\end{figure}
\begin{figure}[h]\center
  \includegraphics[width=.85\linewidth]{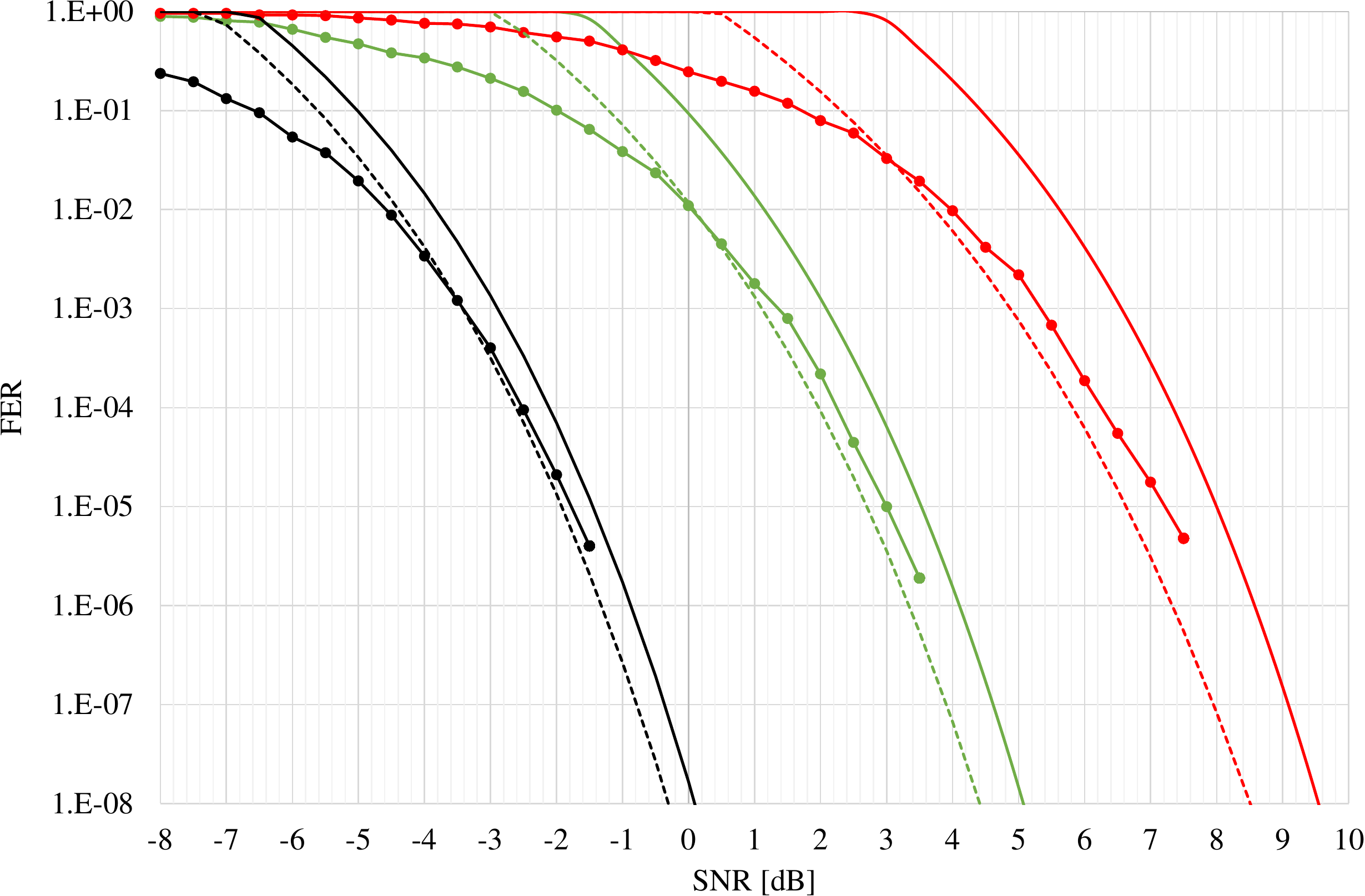}\\
  \caption{Comparison of simulation results and bounds for codes designed for QPSK with non-coherent detection ({\textbf{4RI4}}).
  Target SNR at FER=$10^{-8}$ is 0, 5 and 10 dB, corresponding to required $d_{\mathrm{min}}$ of 43,14, and 5.}\label{fig:QPSKRI}
\end{figure}

To obtain the performance results we used the ML detection rules:
\begin{align*}
y_k&=s_k+n_k             &\mathbf{c}_{ML}^c   =\mathrm{argmax}_\mathbf{\hat{c}}  \Re (\mathbf{y \cdot  s^*(\hat{c})}) &\;\;\mathrm{Coherent}\\
y_k&=s_ke^{j\theta} +n_k &\mathbf{c}_{ML}^{nc}=\mathrm{argmax}_\mathbf{\hat{c}} |\mathbf{y \cdot  s^*(\hat{c})}|      &\;\;\mathrm{Non-Coherent}
\end{align*}

Notice that bound (\ref{eq:ubnc}), is valid only for large values of code length, a situation far from true for the considered short length codes.
This fact is responsible of the mismatch between bounds and simulations observed in Figures \ref{fig:QPSKRI} and \ref{fig:BPSKRI}.

\section{Conclusions and Further Research}\label{sec:conc}
In this paper we presented a strategy to design short-length rate-compatible codes robust to
non-coherent detection. We discussed in some details a system scenario, i.e., the
coding for PLH in precoded satellite broadcast systems, where such codes are necessary to correctly
detect the PLH. A greedy algorithm has been proposed and used to construct the
desired codes. Our method provides an upper bound of the code length needed to assure a given
non-coherent distance between the codewords. The research can be extended in several ways. For
example, it will be interesting to provide some theoretical upper and lower bounds for the
finite-block length coding design with a desired non-coherent distance. Another possible study, is
to change the constellation space instead of using non-coherent detectable codes. As an example, by
adopting a 1-3-APSK constellation one may be able to design shorter codes due to the fact that the
non-coherent detectability as defined earlier may not be needed any more. In general, designing the
constellation space by optimizing an opportunistic objective function (using for example the
techniques proposed in \cite{TWC} and \cite{GB12}) to reduce the code length is an interesting
research topic. Some works in this direction are undergoing.

\section*{Appendix}\label{sec:appendix}

In \cref{tab:tabB,tab:tabQ} we report the generator matrices for the rate-compatible code obtained using the greedy algorithm. The maximum code length
is $N=256$ and the code size is $K=8$. Each rows of the table reports one line of the generating matrix in hexadecimal
notation. Each hexadecimal number has its LSB on the right, so the binary representation should be
read from right to left.

\begin{table}[htbp]
  \centering
 \caption{BPSK codes. Generator matrices of good rate-compatible codes for coherent and non-coherent detection. $K=8$ and maximum code length $N=256$.}
  { \footnotesize \tt
    \begin{tabular}{l|r|r|r|r|r|r|r|r}
       \hline
    &0-31 & 32-63 & 64-95 & 96-127 & 128-159 & 160-191 & 192-223 & 224-255 \\
           \hline
           & \multicolumn{8}{|c}{\rm Coherent detection ($\mathbb{Z}_2$)}  \\
            \hline
   1& B60CC170&	C06D5BD9&	3792D386&	0C019BF6&	A2AC8EA4&	5A6655A6&	0F09AFAF&	A2E06F0C\\
   2& 8740E3C4&	EBF583EA&	40D580B3&	0FDB00DF&	EE798298&	69F03FF3&	3FFFA955&	960395A9\\
   3& 2CB158A6&	3F9EC328&	134ACC79&	C265AB06&	F5554BFF&	95FF363F&	CFFFC965&	F6F0CFF5\\
   4& D7A5A915&	527217DA&	A3FFED21&	E6B15804&	F61AC94F&	F0009C33&	C05C3609&	995C5C96\\
   5& ADD81F34&	A1813BF1&	B187ED2A&	E66B02B1&	65528FD8&	C553CC0C&	9AA553F0&	9AC03F9A\\
   6& CFAABDA7&	C0D26C37&	A2C19ED8&	FF03E402&	1B2CF598&	6F050CA0&	3F0C9F63&	0C9F030C\\
   7& 9E3FFC0F&	9FF5238C&	8BF334D3&	D4EA816C&	8FB01B03&	A60F6CA9&	A6A3CA53&	F74A936A\\
   8& CA4DA8F3&	661EBE67&	CB9558BE&	726B3285&	8B54C955&	0AC0FA09&	036FA3A9&	378306CA\\
            \hline
            & \multicolumn{8}{|c}{\rm Non Coherent detection of BPSK ($\mathbb{Z}_2$)}  \\
            \hline
   -&      FFFFFFFF&FFFFFFFF&	FFFFFFFF&	FFFFFFFF&	FFFFFFFF&	FFFFFFFF&	FFFFFFFF&	FFFFFFFF	\\
    1&     F999750E&F18E7072&	60BB790A&	703FCE66&	ECAABE56&	6096E920&	A0FA9039&	6A5369AA	\\
    2&     6A66FECB&853B0E6B&	3CD19CCF&	733328D4&	188B2829&	6566FB34&	99C5C355&	33659AC0	\\
    3&     650F1B00&F23732A8&	E6B8E6CA&	FC159543&	EC35B180&	FC636E85&	0C93FF09&	A900330A	\\
    4&     CA9BEF81&D397FD46&	64D49C39&	17DA5A4D&	BC102B14&	5999F46D&	555C03C3&	03CA6FA0	\\
    5&     0FF15456&49982A0E&	9F9620F0&	71B330E6&	C7E43FD6&	C03334D8&	5F509663&	3F035A60	\\
    6&     6F602B84&627C1B11&	06D153AC&	17F33F14&	E9594E70&	FF3C6E64&	0A063590&	3A366C0C	\\
    7&     5336D973&CCAE71BF&	45384AF0&	4C3F18C1&	2468DBEA&	953577B0&	FAC6636C&	90A9C59C	\\
    8&     559B1E57&2C354DBA&	94C21C9C&	16BFE8CE&	AF328170&	A0968680&	9003FC36&	F530F995	\\
            \hline
                  \hline
    \end{tabular}%
    }
  \label{tab:tabB}%
\end{table}%

\begin{table}[htbp]
  \centering
 \caption{QPSK codes.Generator matrices of good rate-compatible codes for coherent and non-coherent detection.
 $K=8$ and maximum code length $N=256$.}
  { \footnotesize \tt
    \begin{tabular}{l|c|c|c|c|c|c|c|c}
       \hline
    &0-31 & 32-63 & 64-95 & 96-127 & 128-159 & 160-191 & 192-223 & 224-255 \\
           \hline
           & \multicolumn{8}{|c}{\rm Coherent detection (codes over $\mathbb{Z}_4$)}  \\
            \hline
 1-2&     5CA272B7&	8BB1BEEB&	B645D089&	E8DD708A&	B16F2C92&	BD6C24B3&	390F4A8D&	632DCE31 \\
 3-4&     9E0592A6&	0E8D8D45&	99DDECC8&	3A06F70D&	0866F421&	EE8F409C&	CEE1D497&	E7AD3D2D \\
 5-6&     68B29C05&	9EB80F05&	B6C6B50E&	F443A875&	787406A9&	6C017FD0&	5A4CB78E&	0BBFC41C \\
 7-8&     570C66E7&	2C003E2C&	A05678FA&	5F23FE74&	A8016DB0&	FC76395D&	3E6ED680&	677BAFD8 \\
 \hline
           & \multicolumn{8}{|c}{\rm Non Coherent detection of QPSK (codes over $\mathbb{Z}_4$)}  \\
            \hline
 -&         55555555&	55555555&	55555555&	55555555&	55555555&	55555555&	55555555&	55555555   \\
 1-2&         773F8EF4&	A840C467&	9C9E7E57&	70762CD3&	1214EA38&	AFC0B6CA&	C68A58F0&	FF77C946 \\
 3-4&         EF9E5FF2&	916302C1&	D5D27978&	880B090D&	EF82BA05&	956A5CD2&	5F7DD5D2&	89B64BC5  \\
 5-6&         6CF632A6&	3A8ACF2B&	572D64A5&	5FF77D0D&	AFF031F6&	CD07AF80&	35D00DF0&	E8DFB604  \\
 7-8&         C02A2B67&	839C432D&	3DB1875F&	E7F7C875&	449A58A7&	9FFEBF6B&	4647DD82&	341C8002  \\
           \hline
           & \multicolumn{8}{|c}{\rm Non Coherent detection of QPSK (codes over $\mathbb{Z}_2$)}  \\
            \hline
  -&    FFFFFFFF&	FFFFFFFF&	FFFFFFFF&	FFFFFFFF&	FFFFFFFF&	FFFFFFFF&	FFFFFFFF&	FFFFFFFF  \\
  -&    AAAAAAAA&	AAAAAAAA&	AAAAAAAA&	AAAAAAAA&	AAAAAAAA&	AAAAAAAA&	AAAAAAAA&	AAAAAAAA  \\
  1&    33450EDA&	82F99BB4&	656CA503&	19B61663&	36501FC3&	6A69F93F&	741E26C5&	5A0003E2  \\
  2&    F8252756&	1AD5E6D5&	C939916E&	726C4BCC&	C96ABD0F&	FA000A9C&	4FB93E07&	ABB33000  \\
  3&    9150254A&	B4E249C5&	6A699FEE&	D44BC8A9&	AF30B67B&	E3533006&	6B127392&	0FA0AEBB  \\
  4&    765CC63C&	9C41E7C8&	A5059438&	8EBE5E93&	395C0616&	453FF366&	2E1BAB96&	CE2A3011   \\
  5&    C99A7FB9&	625D30D7&	C09F09AC&	5990AB03&	F96AA05E&	A5636CA3&	5B3BCCB4&	98805987   \\
  6&    FFAA59A6&	A1F655A0&	C3CA37D4&	58D94600&	9A395B36&	7A56AAAA&	83D1A6BB&	CC3C0FD1   \\
  7&    55461830&	B433128F&	5ACA3CE4&	F1E5979A&	59F534D1&	4F5066A0&	E276999B&	9BB991B4   \\
  8&    8F192E8D&	52DB486A&	959A6A0E&	9FF6399A&	C3F9EAD0&	F90F9AC5&	419BB5EF&	3E133CCF    \\
      \hline
    \end{tabular}%
    }
  \label{tab:tabQ}%
\end{table}%

\section*{Acknowledgment}
The authors would like to thank Giulio Colavolpe for several inspiring discussions. Farbod Kayhan
is partially supported by the Luxembourg National Research Fund (FNR) under CORE Junior project:
C16/IS/11332341 Enhanced Signal Space opTImization for satellite comMunication Systems (ESSTIMS).
This research is partially supported by the European Space Agency (ESA) under the TRP contract: ITT
A0/1-8332/15/NL/FE (OPTIMUS). The views of the author do no reflect the views of ESA.

\end{document}